\documentclass[preprintnumbers,showpacs,amsmath,11pt,amssymb,floatfix,superscriptaddress,nofootinbib]{revtex4}
\usepackage{graphicx}
\usepackage{epsfig}
\usepackage{bm}
\usepackage{amsfonts}
\def\be {\begin{equation}}
\def\ee {\end{equation}}
\def\ba {\begin{eqnarray}}
\def\ea {\end{eqnarray}}

\begin{document}
\title{Violation of First Law of Thermodynamics in $f(R,T)$ Gravity}

 \author{\textbf{Mubasher Jamil}}\email{mjamil@camp.nust.edu.pk}
\affiliation{Center for Advanced Mathematics and Physics (CAMP),
National University of Sciences and Technology (NUST), H-12,
Islamabad, Pakistan}\affiliation{Eurasian International Center
for Theoretical Physics, Eurasian National University, Astana
010008, Kazakhstan}

 \author{\textbf{D. Momeni}}
 \email{d.momeni@yahoo.com}
 \affiliation{Eurasian International Center
for Theoretical Physics, Eurasian National University, Astana
010008, Kazakhstan}

 \author{\textbf{Ratbay Myrzakulov}}
 \email{rmyrzakulov@csufresno.edu}
 \affiliation{Eurasian International Center
for Theoretical Physics, Eurasian National University, Astana
010008, Kazakhstan}

\begin{abstract}
\textbf{Abstract:} In this Letter, we derived the first law of
thermodynamics using the method proposed by Wald. Treating the
entropy as Noether charge and comparing with the usual first law of
thermodynamics, we obtained the expression of entropy explicitly
which contains infinitely many non-local terms (i.e. the integral
terms). We have proved, in general, that the first law of black bole
thermodynamics is violated for $f(R,T)$ gravity. But there might
exist some special cases in which the first
law for $f(R,T)$ gravity is recovered. 
\end{abstract}
 \maketitle
\newpage

From the cosmological observational data
\cite{riess,riess2,riess3,riess4}, we know that the present
observable Universe is undergoing an accelerated expansion. While
the source driving this cosmic acceleration is known as `dark
energy' its origin has not been fully understood due to absence of a
consistent theory of quantum gravity. The `cosmological constant' is
the most simple and natural candidate for explaining cosmic
acceleration but it faces serious problems of fine-tuning and large
mismatch between theory and observations \cite{reviews,r2,r3}. Hence
there has been significant development in the construction of dark
energy models by modifying the geometrical part of the
Einstein-Hilbert action. This phenomenological approach is called as
the Modified Gravity which is compatible with the observational data
\cite{NO,noji,carr,j1,j2,j3,Star} (also see a recent review
\cite{review1} on $f(R)$ gravity and its cosmological implications
).

In a recent paper \cite{sergei2011}, the authors considered a
generalized gravity model $f(R,T)$, with $R$ and $T$ being the trace
of Riemann curvature tensor and stress-energy tensor, respectively,
manifesting a coupling between matter and geometry. By choosing
different functional forms of $f$, they solved the dynamical
equations relevant to astrophysical and cosmological interest.
Different aspects of this model discussed in the literatures
\cite{frt1,frt2,frt3,frt4,frt5}. In this paper we discuss the first
law of thermodynamics
 and the Wald's entropy expression for $f(R,T)$ gravity without
 imposing any restrictions on the action. We show that the entropy
 can not be obtained by a simple replacement of the Newtonian gravitational
  constant $8\pi G$ by an effective value $8 \pi G_\text{eff}=8\pi G+f_T$
   (we will set $G=1$ in later calculations) as some authors discussed
   \cite{sharif}. Further, we will show that the first law of the
   thermodynamics is
   violated in $f(R,T)$ gravity. Only when $f_{TT}<<1$,
   the entropy can be obtained using $G_\text{eff}$, but even in this limit
   the first law of the thermodynamics is violated in contradiction to
  previous statements \cite{sharif}.

The action of $f(R,T)$ gravity is given by \cite{sergei2011,epjc}
\begin{equation}\label{1d}
\mathcal{S}=\frac{1}{16\pi}\int d^4x\sqrt{-g}f(R,T) +\int d^4x
\mathcal{L}_m\sqrt{-g},
\end{equation}
where $f(R,T)$ is an arbitrary function of the scalar curvature
$R=R^\mu_\mu$ and the trace $T=T^\mu_\mu$ of the energy-momentum
tensor $T_{\mu\nu}$. This model is an extension of a former model
$f(R)$. In this extension, there is a non-minimally coupling between
the Ricci scalar $R$ and the trace of the energy-momentum tensor. We
define the Lagrangian density for matter field $\mathcal{L}_m$ by
\begin{equation}\label{2d}
T_{\mu\nu}=-\frac{2}{\sqrt{-g}}\frac{\delta(\sqrt{-g}\mathcal{L}_m)}{\delta
g^{\mu\nu}}.
\end{equation}
The equation of motion (EOM) is obtained by varying the action (\ref{1d})
 with respect to $g^{\mu\nu}$ \cite{sergei2011}
\begin{equation}
f_{R}R_{\mu\nu}-\frac{1}{2}fg_{\mu\nu}+(g_{\mu\nu}\Box-\nabla_{\mu}\nabla_{\nu})f_{R}=8\pi
T_{\mu\nu}-f_{T}T_{\mu\nu}-f_{T}\Theta_{\mu\nu}\label{eom1}.
\end{equation}
where $f_R\equiv\frac{\partial f}{\partial R}$ and
$f_T\equiv\frac{\partial f}{\partial T}$. Here $\nabla_\mu$ is the
operator for covariant derivative and box operator (or d' Alembert
operator) $\Box$ is defined via
$$\Box\equiv\frac{1}{\sqrt{-g}}\partial_\mu(\sqrt{-g}g^{\mu\nu}\partial_\nu),\
\ \ \Theta_{\mu\nu}\equiv g^{\alpha\beta}\frac{\delta
T_{\alpha\beta}}{\delta g^{\mu\nu}}.$$ If $f(R,T)$ is replaced with
$f(R)$, this equation is just the $f(R)$ gravity equation. It is
easy to show that the left hand side of this equation is the same as
the $f(R)$ model. The main difference backs to the right hand side.
The contributions from $f_T$ comes in the energy-momentum tensor
part in the right hand side of the gravitational field equation.
Performing a contraction of indices in (\ref{eom1}), we obtain
\begin{eqnarray}\label{eq}
R f_R+3\Box f_R-2f=8\pi T-T f_T-\Theta f_T.
\end{eqnarray}
Here $\Theta\equiv g^{\mu\nu}\Theta_{\mu\nu}$. For a perfect fluid
with the following energy-momentum tensor \be
T_{\mu\nu}=(p+\rho)u_{\mu}u_{\nu}-p g_{\mu\nu}, \ee we have \be
\Theta_{\mu\nu}=-2T_{\mu\nu}-pg_{\mu\nu}. \ee We can write the field
equation for dust $p=0$ in the following form
\begin{equation}\label{eom}
f_{R}R_{\mu\nu}-\frac{1}{2}fg_{\mu\nu}+(g_{\mu\nu}\Box-\nabla_{\mu}\nabla_{\nu})f_{R}=(8\pi+f_T
)T_{\mu\nu}.
\end{equation}
The left hand side of this equation is the same as the equation of
the motion in the $f(R)$ gravity, if $f(R,T)\longrightarrow f(R)$.
So, it is comparable with the $f(R)$ gravity. But if $p\neq0$, then
in the right hand side an extra term $pf_T g_{\mu\nu} $ remains.
This term can be combined in  the left hand side as
$$
-(pf_T+\frac{1}{2}f)g_{\mu\nu}
$$
This term causes some problems in the process of obtaining the
entropy, since now $p(R,T)$ is arbitrary. So it is not clear how we
can treat this term in our calculations. As we know that when the
pressure is included, a \textit{work term} is required to modify the
first law of the thermodynamics. It seems that the physical
interpretation of such term is not easy. Therefore we focus only on
the dust case with $p=0$.

We  adapt Wald's approach to construct entropy and first law of
thermodynamics ($\delta Q=T\delta S$) in any gravity theory
\cite{wald}. Consider a heat flux passing through an open patch
$dH=dAd\lambda$, on a null surface of black hole horizon,
\begin{equation}
\delta Q=\int\limits_H T_{\mu\nu}\xi^\mu k^\nu dAd\lambda,\label{Q}
\end{equation}
where $T_{\mu\nu}$ is the stress energy tensor for energy-matter, $\xi^\mu$ is the killing vector, $H$ represents the null surface, $\lambda$ is the affine parameter, $k^\mu=\frac{dx^\mu}{d\lambda}$ is the tangent vector to $H$. Using (\ref{eom}) in (\ref{Q}), we obtain
\begin{eqnarray}
\delta Q&=&\int\limits_H \frac{1}{8\pi+f_T}[f_R R_{\mu\nu}-\nabla_\mu\nabla_\nu f_R]\xi^\mu k^\nu dAd\lambda,\nonumber\\
&=&\int\limits_H [f_R R_{\mu\nu}-\nabla_\mu\nabla_\nu f_R]\Big(\frac{\xi^\mu}{8\pi+f_T}\Big) k^\nu dAd\lambda,\nonumber\\
&=&\int\limits_H \Big[f_R \nabla_\mu\nabla_\nu \Big(\frac{\xi^\mu}{8\pi+f_T}\Big)-\Big(\frac{\xi^\mu}{8\pi+f_T}\Big)\nabla_\mu\nabla_\nu f_R\Big] k^\nu dAd\lambda,\nonumber
\end{eqnarray}
where we used $R_{\mu\nu}\xi^\mu=\nabla_\mu\nabla_\nu \xi^\mu$ in the above steps. We can further simplify
\begin{eqnarray}
\delta Q&=&\int\limits_H k^\nu\nabla^\mu\Big[ f_R\nabla_\nu\Big( \frac{\xi_\mu}{8\pi+f_T} \Big) \Big]dAd\lambda,\nonumber\\
&=& \int\limits_H k^\nu l^\mu f_R\nabla_\nu\Big( \frac{\xi_\mu}{8\pi+f_T} \Big) dAd\lambda,\nonumber\\
&=& \int\limits_H k^\nu l^\mu f_R\Big[ \frac{\nabla_\nu\xi_\mu}{8\pi+f_T} - \frac{\xi_\mu\nabla_\nu f_T}{8\pi+f_T} \Big] dAd\lambda, \nonumber\\
&=&\frac{1}{8\pi}\int\limits_H k^\nu l^\mu f_R\Big[
\nabla_\nu\xi_\mu \Big( 1-\frac{f_T}{8\pi}+ \Big(\frac{f_T}{8\pi}
\Big)^2-\ldots \Big) - \xi_\mu\nabla_\nu f_T \Big(
1-\frac{f_T}{8\pi}+ \Big(\frac{f_T}{8\pi} \Big)^2-\ldots \Big)\Big]
dAd\lambda,\nonumber
\end{eqnarray}
where we assumed $f_T<1$.  Simplifying further
\begin{eqnarray}
\delta Q&=&\frac{1}{8\pi}\int\limits_H k^\nu l^\mu f_R\nabla_\nu \xi_\mu dAd\lambda-\frac{1}{(8\pi)^2}\int\limits_H k^\nu l^\mu f_Rf_T\nabla_\nu \xi_\mu dAd\lambda+\frac{1}{(8\pi)^3}\int\limits_H k^\nu l^\mu f_Rf_T^2\nabla_\nu \xi_\mu dAd\lambda-\ldots \nonumber\\&&
-\frac{1}{8\pi}\int\limits_H k^\nu l^\mu f_R\xi_\mu\nabla_\nu f_T  dAd\lambda+
\frac{1}{(8\pi)^2}\int\limits_H k^\nu l^\mu f_Rf_T\xi_\mu\nabla_\nu f_T  dAd\lambda
-\frac{1}{(8\pi)^3}\int\limits_H k^\nu l^\mu f_Rf_T^2\xi_\mu\nabla_\nu f_T  dAd\lambda \nonumber\\&&+\ldots \nonumber
\end{eqnarray}

\begin{eqnarray}\label{deltaQ1}
\delta Q&=&\frac{\kappa}{2\pi}\left.\Big(\frac{f_RdA}{4}\Big)\right\vert
_{0}^{d\lambda}-\frac{1}{8\pi}\frac{\kappa}{2\pi}\left.\Big(\frac{f_Rf_TdA}{4}\Big)\right\vert
_{0}^{d\lambda}+\frac{1}{64\pi^2}\frac{\kappa}{2\pi}\left.\Big(\frac{f_Rf_T^2dA}{4}\Big)\right\vert
_{0}^{d\lambda}-\ldots
\\&&
-\frac{1}{8\pi}\int\limits_H k^\nu l^\mu f_R\xi_\mu\nabla_\nu f_T  dAd\lambda+
\frac{1}{(8\pi)^2}\int\limits_H k^\nu l^\mu f_Rf_T\xi_\mu\nabla_\nu f_T  dAd\lambda
-\frac{1}{(8\pi)^3}\int\limits_H k^\nu l^\mu f_Rf_T^2\xi_\mu\nabla_\nu f_T  dAd\lambda \nonumber\\&&+\ldots \nonumber
\end{eqnarray}

It should be stressed that in deriving the first law in
Eq.~(\ref{Q}), we have used the formula
$$
R_{\mu\nu}\xi^\mu=\nabla_\mu\nabla_\nu\xi^\mu
$$
 which is valid only for an exact Killing vector
$\xi^\mu$. However in general for $f(R,T)$, there does not exist any
exact Killing vector in a dynamic spacetime.

It should be mentioned that we have proved that, in general, the
first law of black bole thermodynamics is violated for $f(R,T)$
gravity. But there might exist some special cases in which the first
law for $f(R,T)$ recovers. Note that for black holes with the same
metric $g_{\mu\nu}$, we have many different choices of
energy-momentum tensor $T_{\mu\nu}$ by which they can be related
with each other by some diffeomorphism invariance transformations.
Those black holes have the same $k^{\nu}\nabla_\nu \xi_\mu$ but
different other terms. So, for some special cases of the $f(R,T)$
models, the two terms might cancel each other and the second term of
the last line of Eq.~(\ref{Q}) vanishes.

Now using (\ref{deltaQ1}), we find
\begin{eqnarray}\label{deltaQ}
\delta Q&=&T\delta S-T\delta S_1 +T\delta S_2-\ldots
-\frac{1}{8\pi}\int\limits_H k^\nu l^\mu f_R\xi_\mu\nabla_\nu f_T  dAd\lambda+\ldots
\end{eqnarray}
where
\begin{equation}
\delta S_1= \frac{1}{8\pi}\left.\Big(\frac{f_RdA}{4}\Big)\right\vert
_{0}^{d\lambda},\ \ \delta S_2=\frac{1}{(8\pi)^2}\left.\Big(\frac{f_Rf_TdA}{4}\Big)\right\vert
_{0}^{d\lambda},\ldots\nonumber
\end{equation}
Thus the final expression of entropy in $f(R,T)$ model becomes
\begin{equation}\label{entropy}
S=\frac{f_RA}{4}-\frac{1}{8\pi}\frac{f_Rf_TA}{4}+\frac{1}{64\pi^2}\frac{f_Rf_T^2A}{4} -\frac{1}{4\kappa}\int\limits_H k^\nu l^\mu f_R\xi_\mu\nabla_\nu f_T  dAd\lambda-\ldots
\end{equation}
The last term indicates the entropy production in $f(R,T)$ theory, even in the static cases,
 and this result is the same as the previous result in $f(T)$ gravity \cite{li}.

Since $\nabla_\nu f_T=f_{TT}T_\nu$, and if $f_{TT}$ is small, than (\ref{entropy}) reduces to
\begin{equation}\label{entropy1}
S=\frac{f_RA}{4}-\frac{1}{8\pi}\frac{f_Rf_TA}{4}+\frac{1}{(8\pi)^2}\frac{f_Rf_T^2A}{4} -\ldots
\end{equation}
Alternatively (if $S_{f(R)}\equiv\frac{f_RA}{4}$)
\begin{equation}\label{entropy2}
S=S_{f(R)}-\frac{f_T}{8\pi}S_{f(R)}+\frac{f_T^2}{(8\pi)^2}S_{f(R)} -\ldots
\end{equation}
The first term on right hand side of (\ref{entropy2}) is the Wald
entropy for $f(R)$ gravity which can be obtained by several methods.
But the remaining terms make our results new compared with $f(R)$
gravity. These new terms in entropy arise due to non-minimal
coupling of curvature with matter in the action. It shows that the
thermodynamical aspects $f(R)$ and $f(R,T)$ gravities are completely
different. Further these extra terms modify the Friedmann equations
of $f(R,T)$ gravity.

In this paper, we discussed the thermodynamical properties of
$f(R,T)$ gravity. We derived the first law of thermodynamics using
the method proposed by Wald. Treating the entropy as Noether charge
and comparing with the usual first law of thermodynamics, we
obtained the expression of entropy explicitly which contains
infinitely many non-local terms (i.e. the integral terms) using the
method proposed method by Wald.  Further we corrected some errors
about entropy in a previous work \cite{sharif}. The extra terms in
(\ref{deltaQ}) and (\ref{entropy}) can be understood as `entropy
production' and thus refer to non-equilibrium thermodynamics
\cite{ted}. The first law of thermodynamics does not hold, as a
consequence of entropy production terms in $f(R,T)$ gravity.

\end{document}